# A Cheap Levitating Gas/Load Pipeline

Alexander A. Bolonkin[•]

1310 Avenue R, Suite 6-F

Brooklyn, New York 11229, USA

aBolonkin@juno.com

## Abstract

Design of new cheap aerial pipelines, a large flexible tube deployed at high altitude, for delivery of natural (fuel) gas, water and other payload over a long distance is delineated. The main component of the natural gas is methane which has a specific weight less than air. A lift force of one cubic meter of methane equals approximately 0.5 kg (1 pound). The lightweight film flexible pipeline can be located in air at high altitude and, as such, does not damage the environment. Using the lift force of this pipeline and wing devices payloads of oil, water, or other fluids, or even solids such as coal, cargo, passengers can be delivered cheaply at long distance. This aerial pipeline dramatically decreases the cost and the time of construction relative to conventional pipelines of steel which saves energy and greatly lowers the capital cost of construction.

The article contains a computed project for delivery 24 billion cubic meters of gas and tens of million tons of oil, water or other payload per year.

*Presented in http://arxiv.org, 2008, search "Bolonkin",

## Introduction

Natural gas is a gaseous fossil fuel consisting primarily of methane (CH4) but including significant quantities of ethane, propane, butane, and pentane—heavier hydrocarbons removed prior to use as a consumer fuel —as well as carbon dioxide, nitrogen, helium and hydrogen sulfide. Before natural gas can be used as a fuel, it must undergo extensive processing to remove almost all materials other than methane. The by-products of that processing include ethane, propane, butanes, pentanes and higher molecular weight hydrocarbons, elemental sulfur, and sometimes helium and nitrogen.

Natural gas is not only cheaper, but burns cleaner than other fossil fuels, such as oil and coal, and produces less carbon dioxide per unit energy released. For an equivalent amount of heat, burning natural gas produces about 30% less carbon dioxide than burning petroleum and about 45% less than burning coal.

The major difficulty in the use of natural gas is transportation and storage because of its low density. Natural gas conventional pipelines are economical, but are impractical across oceans. Many existing pipelines in North America are close to reaching their capacity, prompting some politicians representing colder areas to speak publicly of potential shortages.

With 15 nations accounting for 84% of the world-wide production, access to natural gas has become a significant factor in international economics and politics. The world's largest gas field by far is Qatar's offshore North Field, estimated to have 25 trillion cubic meters ($9.0 \times 10^{14}$ cu ft) of gas in place—enough to last more than 200 years at optimum production levels. The second largest natural gas field is the South Pars Gas Field in Iranian waters in the Persian Gulf. Connected to Qatar's North Field, it has estimated reserves of 8 to 14 trillion cubic meters ($2.8 \times 10^{14}$ to $5.0 \times 10^{14}$ cu ft) of gas.

In the past, the natural gas which was recovered in the course of recovering petroleum could not be profitably sold, and was simply burned at the oil field (known as flaring). This wasteful practice is now illegal in many countries. Additionally, companies now recognize that value for the gas may be achieved with LNG, CNG, or other transportation methods to end-users in the future. LNG carriers can be used to transport liquefied natural gas (LNG) across oceans, while tank trucks can carry liquefied or compressed natural gas (CNG) over shorter distances. They may transport natural gas directly to end-users, or to distribution points such as pipelines for further transport. These may have a higher cost, requiring additional facilities for liquefaction or compression at the production point, and then gasification or decompression at end-use facilities or into a pipeline.

Pipelines are generally the most economical way to transport large quantities of oil or natural gas over land. Compared to railroad, they have lower cost per unit and also higher capacity. Although pipelines can be built under the sea, that process is economically and technically demanding, so the majority of oil at sea is transported by tanker ships. The current supertankers include Very Large Crude Carriers and Ultra Large Crude Carriers. Because, when full, some of the large supertankers can dock only in deepwater ports, they are often lightened by transferring the petroleum in small batches to smaller tankers, which then bring it into port. On rivers, barges are often used to transport petroleum.

Pipelines, most commonly transport liquid and gases, but pneumatic tubes that transport solid capsules using compressed air have also been used. Transportation pressure is generally 1,000 pounds per square inch (70 kilograms per square centimeter up to 220 atm) because transportation costs are lowest for pressures in this range. Pipeline diameters for such long-distance transportation have tended to increase from an average of about 24 to 29 inches (60 to 70 centimeters) in 1960 to about 4 feet (1.20 meters). Some projects involve diameters of more than 6 1/2 feet (2 meters). Because of pressure losses, the pressure is boosted every 50 or 60 miles (80 or 100 kilometers) to keep a constant rate of flow.

Oil pipelines are made from steel or plastic tubes with inner diameter typically from 10 to 120 cm (about 4 to 48 inches). Most pipelines are buried at a typical depth of about 1 - 2 metres (about 3 to 6 feet). The oil is kept in motion by pump stations along the pipeline, and usually flows at speed of about 1 to 6 m/s.

For natural gas, pipelines are constructed of carbon steel and varying in size from 2 inches (51 mm) to 56 inches (1,400 mm) in diameter, depending on the type of pipeline. The gas is pressurized by compressor stations and is odorless unless mixed with a mercaptan odorant where required by the proper regulating body. Pumps for liquid pipelines and Compressors for gas pipelines, are located along the line to move the product through the pipeline. The location of these stations is defined by the topography of the terrain, the type of product being transported, or operational conditions of the network.

Block Valve Station is the first line of protection for pipelines. With these valves the operator can isolate any segment of the line for maintenance work or isolate a rupture or leak. Block valve stations are usually located every 20 to 30 miles (48 km), depending on the type of pipeline.

Conventional pipelines can be the target of theft, vandalism, sabotage, or even terrorist attacks. In war, pipelines are often the target of military attacks, as destruction of pipelines can seriously disrupt enemy logistics.

The ground gas and oil pipeline significantly damage the natural environment, but as the demand is so great, ecological concerns are over-ridden by economic factors.

Increased Demand for Natural Gas

Natural gas is a major source of electricity generation through the use of gas turbines and steam turbines. Particularly high efficiencies can be achieved through combining gas turbines with a steam turbine in combined cycle mode. Combined cycle power generation using natural gas is the cleanest source of power available using fossil fuels, and this technology is widely used wherever gas can be obtained at a reasonable cost. Fuel cell technology may eventually provide cleaner options for converting natural gas into electricity, but as yet it is not price-competitive.

Natural gas is supplied to homes, where it is used for such purposes as cooking in natural gas-powered ranges and/or ovens, natural gas-heated clothes dryers, heating/cooling and central heating. Home or other building heating may include boilers, furnaces, and water heaters. CNG is used in rural homes without connections to piped-in public utility services, or with portable grills. However, due to CNG being less economical than LPG, LPG (Propane) is the dominant source of rural gas.

Compressed natural gas (methane) is a cleaner alternative to other automobile fuels such as gasoline (petrol) and diesel. As of 2005, the countries with the largest number of natural gas vehicles were Argentina, Brazil, Pakistan, Italy, Iran, and the United States. The energy efficiency is generally equal to that of gasoline engines, but lower compared with modern diesel engines. Gasoline/petrol vehicles converted to run on natural gas suffer because of the low compression ratio of their engines, resulting in a cropping of delivered power while running on natural gas (10%-15%). CNG-specific engines, however, use a higher compression ratio due to this fuel's higher octane number of 120-130.

Russian aircraft manufacturer Tupolev is currently running a development program to produce LNG- and hydrogen-powered aircraft. The program has been running since the mid-1970s, and seeks to develop LNG and hydrogen variants of the Tu-204 and Tu-334 passenger aircraft, and also the Tu-330 cargo aircraft. It claims that at current market prices, an LNG-powered aircraft would cost 5,000 roubles (~ $218/ £112) less to

operate per ton, roughly equivalent to 60%, with considerable reductions to carbon monoxide, hydrocarbon and nitrogen oxide emissions.

The advantages of liquid methane as a jet engine fuel are that it has more specific energy than the standard kerosene mixes and that its low temperature can help cool the air which the engine compresses for greater volumetric efficiency, in effect replacing an intercooler. Alternatively, it can be used to lower the temperature of the exhaust.

Natural gas can be used to produce hydrogen, with one common method being the hydrogen reformer. Hydrogen has various applications: it is a primary feedstock for the chemical industry, a hydrogenating agent, an important commodity for oil refineries, and a fuel source in hydrogen vehicles. Natural gas is also used in the manufacture of fabrics, glass, steel, plastics, paint, and other products.

It is difficult to evaluate the cost of heating a home with natural gas compared to that of heating oil, because of differences of energy conversion efficiency, and the widely fluctuating price of crude oil. However, for illustration, one can calculate a representative cost per BTU. Assuming the following current values (2008):
For natural gas.

- One cubic foot of natural gas produces about 1,030 BTU (38.4 MJ/m³).

- The price of natural gas is $9.00 per thousand cubic feet ($0.32/m³).

For heating oil.

- One US gallon of heating oil produces about 138,500 BTU (38.6 MJ/l).

- The price of heating oil is $2.50 per US gallon ($0.66/l) .

This gives a cost of $8.70 per million BTU ($8.30/GJ) for natural gas, as compared to $18 per million BTU ($17/GJ) for fuel oil. Of course, such comparisons fluctuate with time and vary from place to place dependent on the cost of the raw materials and local taxation.

### *Current pipelines.*

**Natural gas pipelines**. The long-distance transportation of natural gas became practical in the late 1920s with improvements in pipeline technology. From 1927 to 1931 more than ten major gas pipeline systems were built in the United States. Gas pipelines in Canada connect gas fields in western provinces to major eastern cities. One of the longest gas pipelines in the world is the Northern Lights pipeline, which is 3,400 miles (5,470 kilometers) long and links the West Siberian gas fields on the Arctic Circle with locations in Eastern Europe.

**Oil and petroleum products pipelines**. Pipelines are used extensively in petroleum handling. In the field, pipes called gathering lines carry the crude oil from the wells to large storage depots located near the oil field. From these depots the oil enters long-distance trunk lines, which may carry it to an intermediate storage point or directly to refineries.

By the last quarter of the 20th century, there were about 250,000 miles (400,000 kilometers) of oil pipeline in operation in the United States. About one third of the total mileage consisted of crude-oil trunk lines. Pipelines carry large volumes of crude oil from fields in Texas, Louisiana, and Oklahoma to refineries in the Midwest and on the East coast of the United States. The 800-mile (1,300-kilometer) north-south trans-Alaskan oil pipeline, which began operation in 1977, connects the Prudhoe Bay fields on the northern coast of Alaska to Valdez on Prince William Sound. Europe also has several crude-oil pipelines supplying inland refineries. In the Middle East large pipelines carry crude oil from oil fields in Iraq, Saudi Arabia, and other oil-exporting countries to deepwater terminals on the Mediterranean Sea.

From the refinery, large volumes of petroleum products travel to the market area by way of product lines. In areas of high consumption, several similar products gasoline, furnace oil, and diesel oil, for example may be shipped in the same line in successive batches of several tens or hundreds of thousands of barrels each. The first European product line, called the TRAPIL line, was completed in 1953 in France. In 1964 the world's largest products line began operation in the United States; the pipeline can transport 1 million barrels of products per day from Houston, Tex., to New Jersey.[1]

**The trans-Alaskan pipeline** was the most expensive pipeline in the world costing 9 billion dollars to build. It is 48 inches (122 centimeters) in diameter and 800 miles (1,300 kilometers) long. Oil moves southward through it, at a rate of 1.5 million barrels each day, from the giant Prudhoe Bay oil field on the northern coast of Alaska to the ice-free port of Valdez, where the oil is shipped by tankers to refineries on the West coast of the United States. The pipeline traverses three mountain ranges and 250 rivers and streams. For some 400 miles (640 kilometers) it is suspended on pylons above permanently frozen ground called permafrost. If the pipeline had been buried in the ground, the heat from the oil in the pipeline would have melted the permafrost, causing considerable environmental damage.

This is a list of countries by total length of pipelines mostly based on The World Fact book  accessed in June 2008 .

| Rank | Country | Total length of pipelines (km) | Structure | Date of Information |
|---|---|---|---|---|
| 1 | United States | 793,285 | petroleum products 244,620 km; natural gas 548,665 km | 2006 |
| 2 | Russia | 244,826 | condensate 122 km; gas 158,699 km; oil 72,347 km; refined products 13,658 km | 2007 |
| 3 | Canada | 98,544 | crude and refined oil 23,564 km; liquid petroleum gas 74,980 km | 2006 |
| 4 | China | 49,690 | gas 26,344 km; oil 17,240 km; refined products 6,106 km | 2007 |

---

[1] *Reference:* Compton's Interactive Encyclopedia.

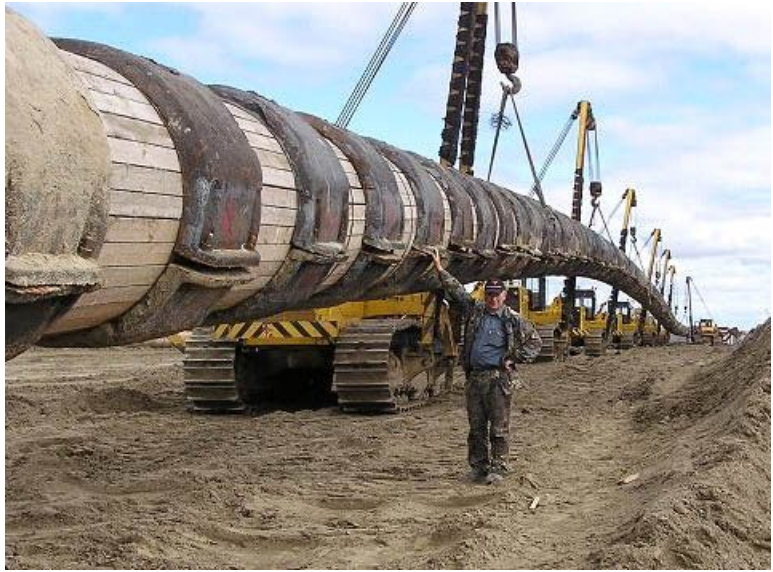

Building typical ground pipeline.
Requires ground right of way and results in damage to ecology

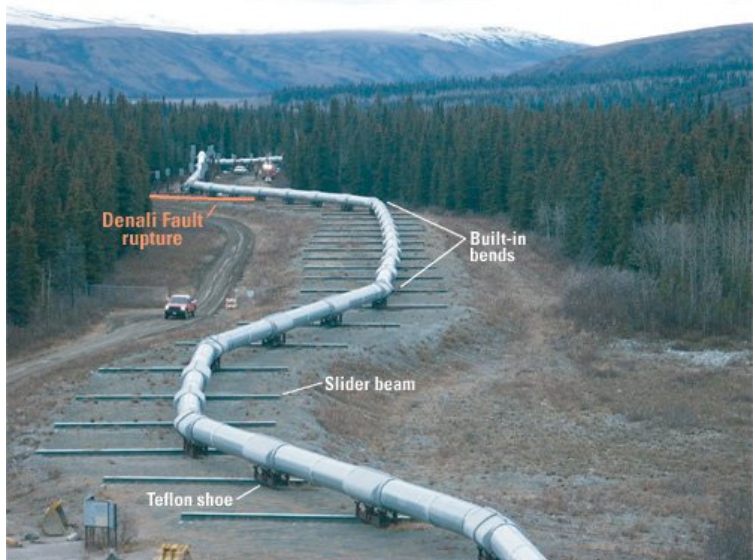

Trans-Alaska oil pipeline.
Notice the damage.

*Some currently planned pipeline projects*:

The **Nabucco pipeline** is a planned natural gas pipeline that will transport natural gas from Turkey to Austria, via Bulgaria, Romania, and Hungary. It will run from Erzurum in Turkey to Baumgarten an der March, a major natural gas hub in Austria. This pipeline is a diversion from the current methods of importing natural gas solely from Russia which exposes EC to dependence and insecurity of the Kremlin practices. The project is backed by the European Union and the United States.

The pipeline will run from Erzurum in Turkey to Baumgarten an der March in Austria with total length of 3,300 kilometers (2,050 mi). It will be connected near Erzurum with the Tabriz-Erzurum pipeline, and with the South Caucasus Pipeline, connecting Nabucco Pipeline with the planned Trans-Caspian Gas Pipeline. Polish gas company PGNiG is studying the possibility of building a link to Poland with the Nabucco gas pipeline.

In the first years after completion the deliveries are expected to be between 4.5 and 13 billion cubic meters (bcm) per annum, of which 2 to 8 bcm goes to Baumgarten. Later, approximately half of the capacity is expected to be delivered to Baumgarten and half of the natural gas is to serve the markets en-route. The transmission volume of around 2020 is expected to reach 31 bcm per annum, of which up to 16 bcm goes to Baumgarten. The diameter of the pipeline would be 56 inches (1,420 mm).

The project is developed by the Nabucco Gas Pipeline International GmbH. The managing director of the company is Reinhardt Mitschek. The shareholders of the company are: OMV (Austria), MOL (Hungary), Transgaz (Romania), Bulgargaz, (Bulgaria), BOTAŞ (Turkey), RWE (Germany).

In 2006, Gazprom proposed an alternative project competing Nabucco Pipeline by constructing a second section of the Blue Stream pipeline beneath the Black Sea to Turkey, and extending this up through Bulgaria, Serbia and Croatia to western Hungary. In 2007, the South Stream project through Bulgaria, Serbia and Hungary to Austria was proposed. It is seen as a rival to the Nabucco pipeline. Ukraine proposed White Stream, connecting Georgia to Ukrainian gas transport network.[2]

These mega-pipeline projects and others currently planned will require investment of at lease $200Billion in the next few years. We propose a much cheaper alternative, an aerial pipeline that is lifted because natural gas is lighter than air and as such a lifting gas. An inflatable pipeline pressurized with natural gas will levitate and float up as if full of helium. Each end of the pipeline will be tethered to the ground but the middle will soar above land and sea at altitudes between .1 and 6 kilometers. It can span seas and can be up to a hundred times cheaper than conventional undersea pipelines. It can be only 100meters above the ground and easily monitored and repaired.

The main differences of the suggested Gas Transportation Method and Installation from current pipelines are:

1. The tubes are made from a lightweight flexible thin film (no steel or solid rigid hard material).
2. The gas pressure into the film tube equals an atmospheric pressure or less more (<1.5 atm.) (The current gas pipelines have pressure up 220 atmospheres.).
3. The great majority of the pipeline [except compressor (pumping) and driver stations] is located in atmosphere at a high altitude (0.1-6 km) and does not have a rigid compression support (pillar, pylon, tower, bearer, etc.) though it may be tethered. The current pipelines are located on the ground, under ground or under water.

---

[2] http://en.wikipedia.org/wiki/Category:Future_pipelines.

4. The transported gas contained within the pipeline supports the gas pipeline at high altitude.
5. Wing devices may make additional lift.
6. The gas pipeline can be used as an air transport system for oil and solid payloads with a high speed for each container shipped up of 250 m/sec.
7. The gas pipeline can be used as for a large scale transfer of mechanical energy.

The suggested Method and Installation have huge advantages in comparison with modern steel gas pipelines.

## Description of innovation

A gas and payload delivering gas/load pipeline is shown on figs. 1 - 6.
Fig.1 shows the floating pipeline.

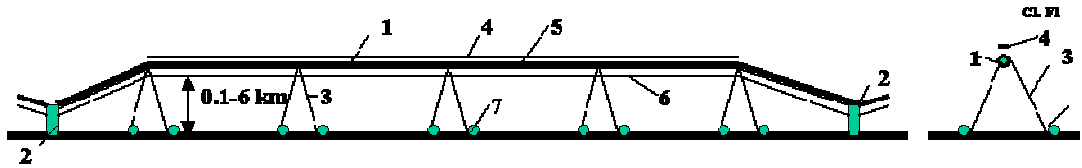

**Fig.1.** General view of air pipeline between two compression (pump) stations. (*a, left*) Side view, (*b-right*) front view. Notations: 1 - pipeline in atmosphere, 2 - compression station, 3 - tensile stress element, 4 - support wing device, 5 – aircraft warning light, 6 - load (container) monorail, 7 - winch.

The installation works as follows: The compressor station pumps in gas from the system input (natural gas processing plant, storage system, etc) into the floating pipeline. The tube is made from light strong flexible gas-impermeable fire-resistant material (film), for example, from composite material (fibers, whiskers, nanotubes, etc.). The internal gas pressure is a bit higher than the outside atmospheric pressure (up 0.1- 0.2 atm) both to keep the tube inflated and to send the contents flowing toward the far end of the pipeline. Natural gas has methane as a main component with a specific density about 0.72 kg/m$^3$. Air, by contrast, has a specific density about 1.225 kg/cubic meters. It means that every cubic meter of gas (methane) or a gas mixture with similar density  has a lift force approximating 0.5 kg. Each linear (one meter) weight of the tube is less than the linear lift force of the gas in the tube and the pipeline thus has a net lift force. The pipeline rises up and is held at a given altitude (0.1 - 6 km) by the tensile elements 3. The altitude can be changed by the winches 7. The compressor station is located on the ground and moves the flowing gas to the next compressor station which is usually located at the distance 70 - 250 km from previous. Inside of the pipeline there are valves (fig.4) dedicated to lock the tube in case of extreme damage. The pipeline has also the warning light 5 for aircraft.

Fig.2 shows the cross-section of the gas pipeline and support ring.  The light rigid tube support ring helps intermediate the motions and stresses of, variously, the lift force from the gas tube, wing support devices, from monorail and load containers, and the tension elements (tiedown tethers) 5.

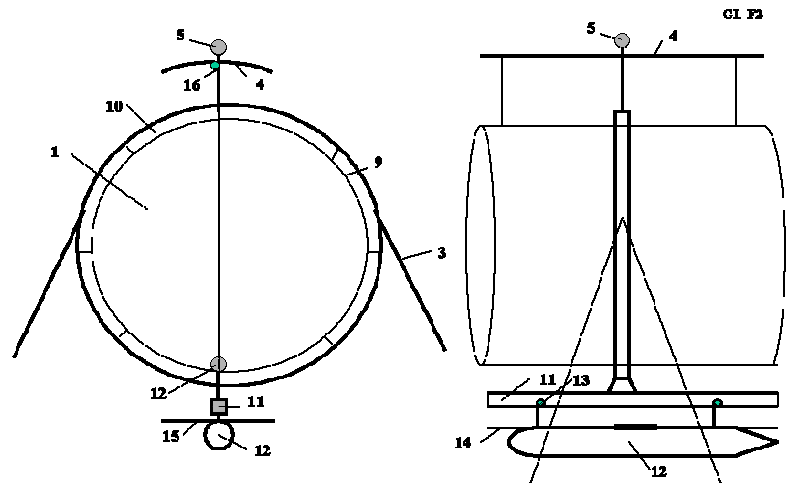

**Fig.2.** Cross-section and support ring of aerial pipeline. (*a-left*) front view, (*b-right*) side view. Notations: 9 - double casing, 10 -rigid ring, 11 - monorail, 12 - wing load container, 13 - rollers of a load container, 14 - thrust cable of container, 15 - container wing. Other notations are same as in fig.1.

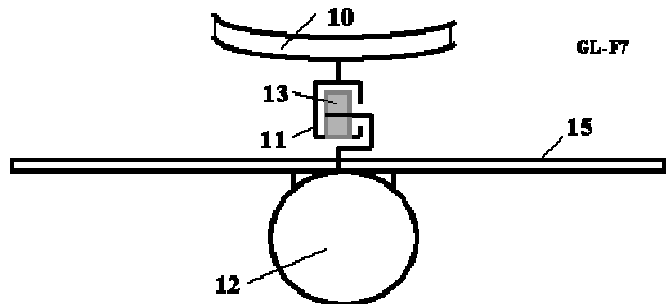

**Fig, 2a**. Front view of the wing load container, monorail, and suspension bracket. Other notations are same as    figs.1-2.

Fig.3 shows the compressor (pumping) station. The station is located on the ground or water and works in the following way. The engine 32 rotates compressor 31. This may be in the form of propellers located inside a rigid ring body connected to the flexible tubes of installation 1. The pipeline tubes are located floating in atmosphere near the compressor. The propeller moves the gas in the given direction.

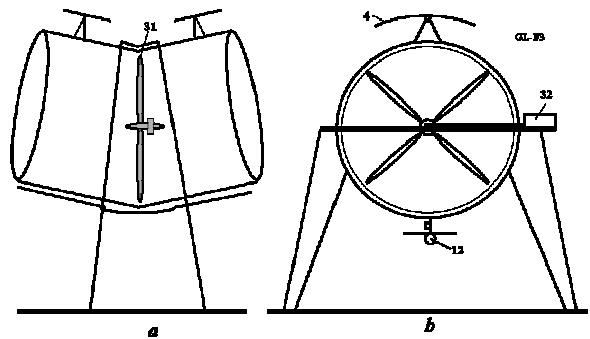

**Fig.3.** Compression (pumping) station. Notations: 31 - compressor (propeller), 32 - engine. Other notations are same as figs.1-3.

Fig.4 shows two variants of a gas valve. The first valve version is an inflatable ball. The ball would expand out and close the gas path. The second version is a conventional light flat valve-closing disc in a  circular form.

The valve works in the following way. When the tube section is damaged, the pressure decreases in a given section. The control devices measure it, the valves close the pipeline. The valve control devices have a connection with a central dispatcher, who can close or open any section of the pipeline.

Fig.4a shows the spherical valve (a ball) in packed (closed) form. Fig.4b shows the spherical valve (a ball) in an open form.

The tubes of pipeline have a double wall (films) and gas sealing liquid may be between them. If the walls damage the streak gas flows out and the second film closes the hole in the first film and conserves the pipeline's gas.

The wing device 4 is a special automatic wing. When there is a wind whether and the side wind produce a strong side force, the wing devices produce a strong lift force and supports the pipeline in a vertical position.

The wind device works in the following way. When there is a side wind the tube has the wind drug and the wind device creates the additional lift force. All forces (lifts, drags, weights) are maintained toward equilibrium. The distance between the tensile elements 3 is such that the tube can withstand the maximum storm wind speed. The system can have a compensation ring. The compensation ring includes ring, elastic element, and cover. The ring compensates the temperature's change of the tube and decreases stress from wind.

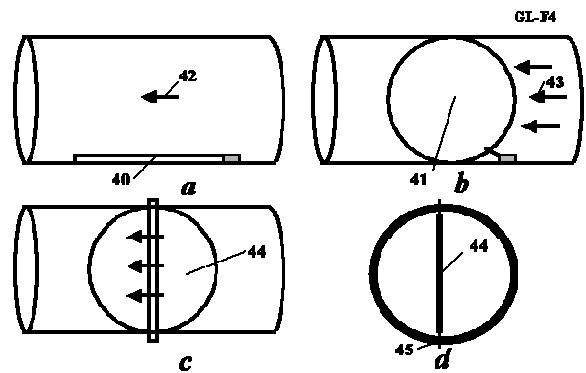

**Fig. 4**. Gas valve. (a)-(b) Inflatable valve, (c)-(d)  Flat valve. Notations: 40 - inflatable valve in compact position, 41 - inflatable valve in fill out position, 42 - gas flow, 43 - gas pressure, 44 - flat choker disc.

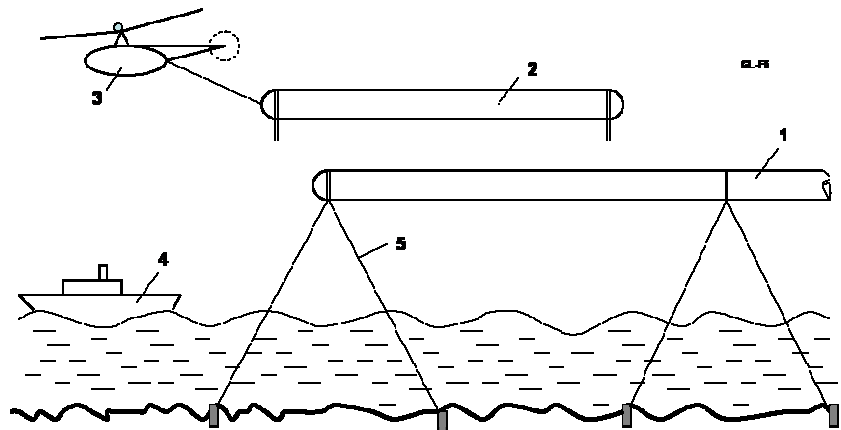

**Fig.5**. Building the offered gas pipeline. Notation: 1 – gas pipeline; 2 – delivered section of gas pipeline; 3 – deliver helicopter; 4 – support ship; 5 – tensile links.

The suggested gas pipeline has advantages over the conventional steel gas pipeline:

1. The suggested gas pipeline is made from a thin film is hundreds times less expensive than a current gas pipeline made from ground or sea-bottom bound steel tubes.
2. The construction time may be decreased from years to months.
3. There is no need to supercompress a gas (over 200 atmospheres) and spend huge energy for that.
4. No large area of expensive user rights needed on ground surface and little surface environmental damage during the building and use of the pipeline.
5. No surface environmental damage in case of pipeline damage during use of the pipeline.
6. Easy to repair.
7. Less energy used in gas transit and delivery.
8. Additional possibility of payload delivery in both directions.
9. If the pipeline is located at high altitude, it is more difficult for terrorist diversion and for petty or large scale gas stealing by third parties when the suggested pipeline cross the third country (for example, unknown parties in the Ukraine typically are said to steal 10-15% amount of the gas from Russian pipelines transiting the Ukraine.).
10. The suggested transportation system may be also used for a transfer of mechanical energy from one driver station to another, more cheaply than electric power could be sent.

More detailed description of the innovation may be found in publications [1]-[8]. Below, projects suitable for many regions are proposed. This paper is by no means exhaustive and apparent problems in this project, such as standing up to strong winds, have been solved and the author is prepared to discuss the problems and solutions with serious organizations, which are interested in researching and developing these ideas and related projects.

## Methods of the estimation of the altitude gas pipeline

1. Gas **delivery capability** is
$$G = \pi D^2 V/4 \ \text{[m}^3\text{/sec]}, \quad M = \rho G \ \text{[kg/s]},\qquad(1)$$

where: $D$ is diameter of tube [m]; $V$ is average gas speed [m/sec], $M$ is mass flow [kg/s], $\rho = 0.74$ kg/m$^3$ is gas methane density.  Result of computation is presented in Fig. 6.

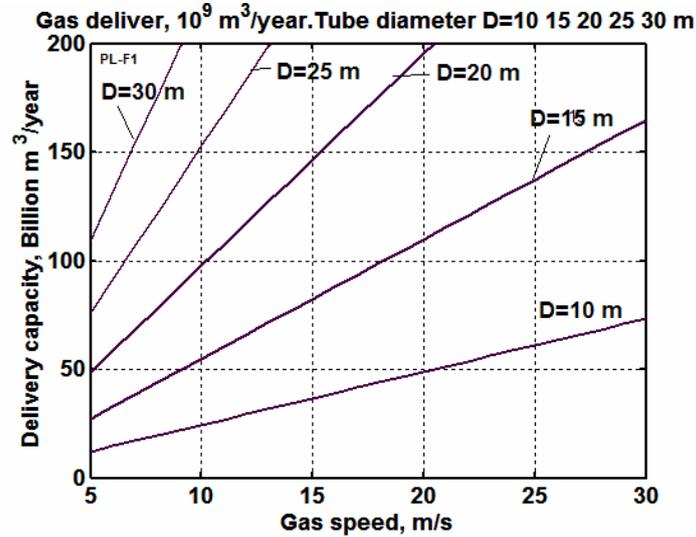

**Fig. 6**. Delivery capacity of gas by the offered floating gas pipeline by gas speed and tube diameter.

2. Increment **pressure** [N/m$^2$] is

$$P = \lambda L \rho V^2 / 2D \quad , \tag{2}$$

where: $\lambda$ is dimensionless factor depending on the wall roughness, the fluid properties, on the velocity, and pipe diameter ($\lambda = 0.01 - 0.06$); $L$ is the distance between pump station [m]; $\rho$ is fluid density [kg/m$^3$].

The dimensionless factor can be taken from graph [5, p.624]. It is

$$\lambda = f(R, \varepsilon) \quad , \tag{3}$$

where: $R = VD\rho/\mu$ is Reynolds number, $\mu$ is fluid viscosity; $\varepsilon$ is measure of the absolute roughness of tube wall. In our case $\lambda = 0.015$.

Result of computation of the gas pressure in relation to the distance between pump station and gas speed are presented in Fig. 7.

3. **Lift force** $F$ of each one meter length of pipeline is

$$F = (\upsilon_a - \upsilon_m)\boldsymbol{v} \quad , \tag{4}$$

where: $\upsilon_a = 1.225$ kg/m$^3$ is air density for standard atmosphere; $\upsilon_m = 0.72$ kg/m$^3$ is methane density; $v = \pi D^2/4$ is volume of one meter length of pipeline, $F$ [kg/m]. Results of computation are presented in Fig.8.

4. Needed **thickness** $\delta$ of tube wall is

$$\delta = PD/2\sigma \quad , \tag{5}$$

where: $P$ is pressure [see Eq.(2)]; $\sigma$ is safety stress. That is equal to 50 - 200 kg/mm$^2$ for materials made from current artificial fibers. (Stronger materials, of course, would allow better results) Results of computation are presented in Fig.9.

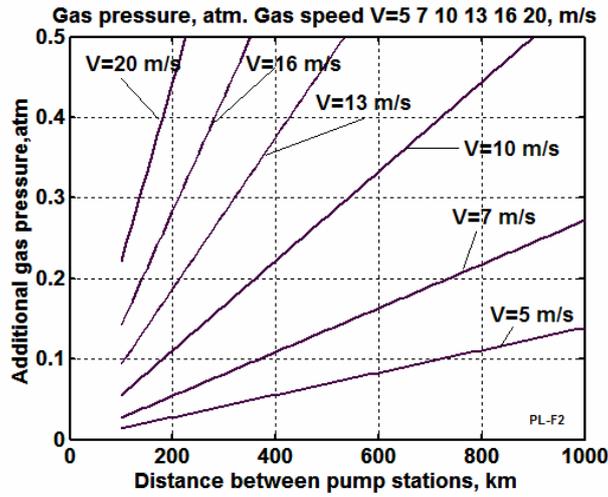

**Fig.7.** Gas pressure versus the distance between pump stations for different gas speeds.

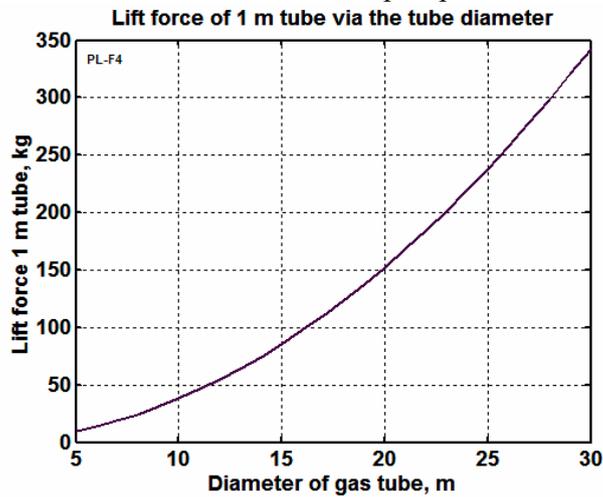

**Fig.8.** Lift force of each1 meter of methane gas bearing tube in relation to tube diameter.

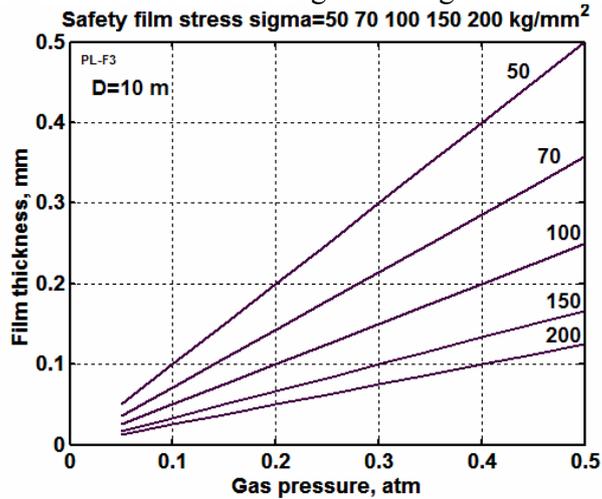

**Fig. 9.** Film thickness via gas pressure and the safety film stress for the tube diameter $D = 10$ m.

5. **Weight** of one meter of pipeline is

$$W = \pi D \delta \gamma \ , \tag{6}$$

where: $\gamma$ is specific weight of matter tube material (film, cover). That equals about $\gamma = 900 - 2200$ kg/m$^3$ for matter from artificial fiber.

6. **Air drag** $D$ of pipeline from side wind is

$$D = C_d \rho V^2 S/2 \ , \tag{7}$$

where: $C_d = 0.01 \sim 0.2$ is drag coefficient; $S$ is logistical pipeline area between tensile elements; $\rho = 1.225$ kg/m$^3$ is air density.

7. Needed **pumping power** is

$$N = PG/\eta \ , \tag{8}$$

where $\eta \approx 0.9$ is the efficiency coefficient of a compressor station. Result of computation is presented in Fig.10.

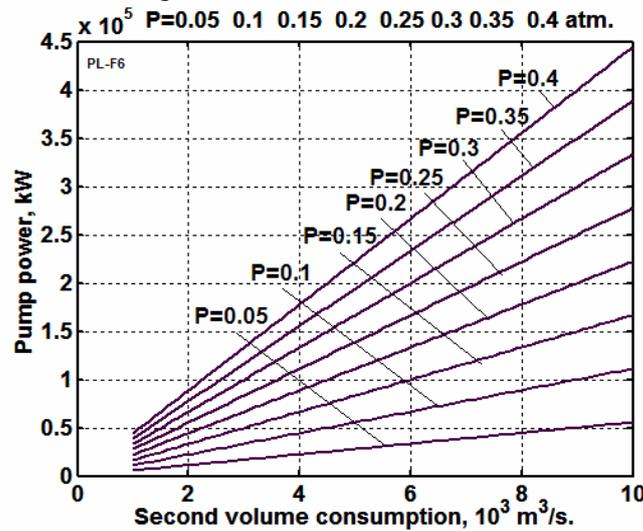

**Fig.10**. Pumping power versus the volume consumption and gas pressure.

We may also view that pumping power equals

$$N = 2C_f M V^2 L/D\eta, \tag{8a}$$

where $M$ is mass of the gas [kg/s], $L$ is length of pipeline between compression station [m], $C_f$ is the friction coefficient. That means the delivery power decreases when we use the tubes of bigger diameter. In our case the chosen diameter is 10-20 m. That means the pumping power (to force the gas through the pipeline) will be about 10 times less than in a conventional high pressure pipeline having maximal diameter of 1.44 meters. The conventional steel pipeline method requests also a lot of additional pumping energy for gas compression of up to 70 – 200 atmospheres.

**Load container transportation system mounted under pipeline.**

1. **Load delivery** capability by wingless container is

$$G_p = kFV_p \ , \tag{9}$$

where: $k$ is load coefficient $(k \approx 0.5 < 1)$; $V_p$ is speed of container (load). Result of computation is presented in Fig.11.

2. **Friction force** of wingless containers (using wheels or rollers on a monorail) is

$$F_c = f W_c \ , \tag{10}$$

where: $f \approx 0.03 - 0.05$ is coefficient of roller friction; $W_c$ is weight of containers between driver stations.

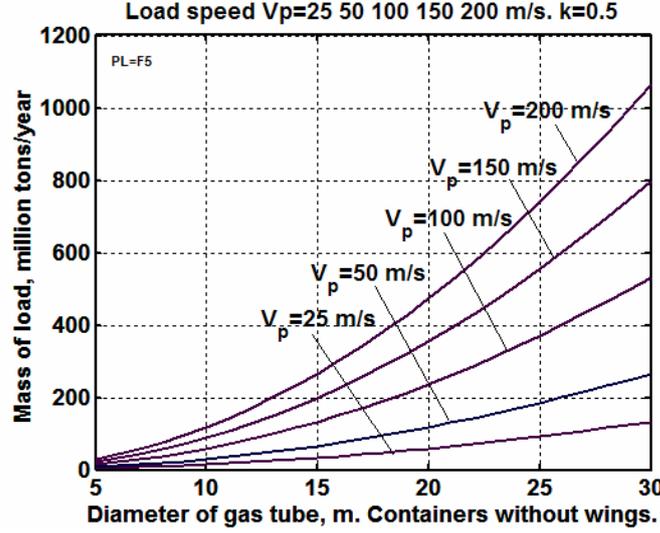

**Fig.11**. Load delivery via the tube diameter and speed payload by wingless containers for $k$=0.5. Delivery by winged container may be some times more. The load delivery of a conventional two-line railroad is about only 15-30 million tons/year. But a railroad's capital cost is more expensive by many times than the offered high-speed aerial transport system. The big (diameter 1.4 m) oil pipeline has delivery capability of about 100 million tons/year.

3.  **Air drag of a container** is

$$D_c = C_c \rho V^2 S_c / 2 \ , \tag{11}$$

where: $C_c$ is drug friction coefficient related to $S_c$; $S_c$ is the cross-sectional area of container.

4.  The **lift force** of a winged container is

$$L_c = C_L q S_{cw} = C_L \frac{\rho_a V^2}{2} S_{cw} , \tag{12}$$

where: $C_L \approx 1 - 1.5$ is coefficient of lift force; $q = \rho_a V^2 / 2$ is air dynamic pressure, N/m$^2$; $S_{cw}$ is wing area of winged container, m$^2$.

5.  The **drag of a winged container** may be computed by equation

$$D_C = C_L q S_{cw} / K = C_L \frac{\rho_a V^2}{2K} S_{cw} , \quad \text{where} \quad K = \frac{C_L}{C_D} , \tag{13}$$

where $K \approx 10 - 20$ is the coefficient of aerodynamic efficiency; $C_D$ is air drag coefficient of winged container. If lift force of wing container equals the container weight, the friction force $F$ is absent and in theory it does not require a solid monorail.

6. The **delivery (load) capacity** of the winged container system as a whole is

$$G_c = \frac{W_1 V_c T}{d} , \tag{14}$$

where $W_1$ is weight of one container, kg; $V_c$ = 30 - 200 m/s is container speed, m/s; $T$ is time, s; $d$ is distance between two containers, in meters.

7. The **lift and drag** of the wing device may be computed by Equations (12)-(13). The power needs for transportation system of winged containers is found by

$$P_c = \frac{gWV_c}{K_c},\qquad (15)$$

where $W$ is total weight of containers, kg; $g = 9.81$ m/s$^2$ is Earth gravity; $K_c \approx 10$ - 20 is aerodynamic efficiency coefficient of container and thrust cable;

8. The **stability** of the pipeline against a side storm wind may be estimated by the inequality

$$A = \frac{L_T + L_d - gW_T - gW_S}{D_T + L_d / K_d} > \tan \alpha > 0,\qquad (16)$$

where $L_T$ is lift force of given part of pipe line, N; $L_d$ is lift force of wing device, N; $W_T$ is a weight of pipeline of given part, kg; $W_s$ is weight of the given part suspending system (containers, monorail, thrust cable, tensile element, rigid ring, etc.), kg; $D_T$ is drag of the given part of pipeline, N; $L_d$ is the lift force of the wing device, N; $K_d$ is an aerodynamic efficiency coefficient of wing device; $\alpha$ is the angle between tensile element and ground surface.

# Project

*(Tube diameter equals D = 10 m, gas pipeline has the suspended- load transport system, the project is suitable for many regions)*

Let us take the tube diameter $D = 10$ m, the distance between the compressor-driver stations $L = 100$ - 1000 km and a gas speed $V = 10$ m/sec.

Gas delivery capacity is then (Eq. (1))

$$G = \pi D^2 V/4 = 800 \text{ m}^3\text{/s} = 25 \text{ billions m}^3 \text{ per year [km}^3\text{/year]}.$$

For the Reynolds number $R = 10^7$ value $\lambda$ is 0.015, $P = 5 \times 10^3$ N/m$^2$ = 0.05 atm (Eq. (2)-(3) for $L = 100$ km). We can take V = 20 m/s and decrease delivery capacity in two (or more) times.

Lift force (Eq. 4) of one meter pipeline's length equals $F = 39$ kg for diameter $D = 10$ m.. For $\sigma = 200$ kg/sq.mm the tube wall thickness equals $\delta = 0.0125$ mm (Eq. 5 for $P = 0.05$ ). We take the thickness of wall as $\delta = 0.15$ mm. That is enough **for a distance of 1200 km between pumping stations!**

The cover weight of each one meter of the pipeline's length is 7 kg [Eq. (6)]. The needed power of the compressor station (located at distance of 1000 km) equals $N = 44,400$ kW for $\eta = 0.9$ [Eq.(8)].

## Cost of altitude gas pipeline

Let us take the film thickness $\delta = 0.15$ mm, specific weight of film $\gamma = 1440$ kg/m3, film cost $c = \$4$/kg, diameter of tube $D = 10$ m. Then volume of 1 m of tube is $v = \pi D\delta = 4.7 \times 10^{-3}$ m$^3$, weight is $W = \gamma v = 6.77$ kg/m, cost of 1 m tube film is $C_1 = cW = \$28$/m.

The tube needs one wing attached to the tube for each length 10 m. Let us take the cost of one such wing \$20. It is $C_2 = \$2$/m. Support cables and anchors for each tube length of 100 m cost about \$200 or $C_3 = \$2$/m. Let us take the cost of pumping station good for 100 km is \$100,000 and good for 1000 km the \$1 mln. That means the cost of the pumping station is $C_4 = \$1$/m.

The total cost of 1 m of pipeline (material) is $C = C_1 + C_2 + C_3 + C_4 =$ \$33/m or $C = 33K/km$ (K is thousand). The Labor and additional works usually costs 100-200% from material. That means each 1 km of the offered altitude pipeline may cost as little as \$100K/km after the tooling and other overhead costs are paid for.

In conventional pipelines the steel tubes alone cost $0.3 - 1$ million dollars per km. The builder must also pay a lot of money for the right of way on the Earth's surface. The deliver must also annually pay fees for gas transition through the territory of other countries - transition rights. The offered pipelines are very cheap and in many cases may be located over the high seas, cutting down greatly on the right of way fees.

Compare the cost of the current and proposed ground and sea pipelines in Table 1. As you see the offered pipe is $\sim 10 - 100$ times cheaper than conventional ground pipelines of same length. They may be built if need be more rapidly; in a year or so, with extra money to cover the expense of such rapid construction, in as little as 5-6 months, to be contrasted with 4-5 years as conventional pipelines. This time value of capital saved is a major advantage.

Such a pipeline does not require payment for transition across other countries via ground rights (air rights are usually far cheaper because the uses of the land may continue below) and they require 10 - 15 times less energy for pumping gas because there is no great compression and less gas friction.

**Table 1.** Design and current pipelines.

| # | Pipeline | Length, km | Sea part, km | Cost 1 line, $Billion | Cost 1 km, $mln/km | Min. diameter, m | Deliver capacity, km³/year | Final year of building |
|---|----------|-----------|--------------|----------------------|-------------------|------------------|---------------------------|------------------------|
| 1 | Russia-German | 1200 | 1200 | 5 | 4.2 | 1.44 | 27.5 | 2010 |
| 2 | Russia-Bolgaria | 900 | 900 | 10 | 11 | ? | 30 | 2013 |
| 3 | Asia-Europa | 3300 | - | 10 | 3 | 1.42 | 4.5-28 | 2013 |
| 4 | Russia-Turkey | 1230 | 396 | 3.2 | 2.64 | 0.61 | 16 | 1997 |
| 5 | Albania-Italy | 520 | - | 1.9 | 3.65 | ? | 10 | 2011 |
| 6 | Turkey-Greece | 296 | 17 | ? | - | 0.914 | 7 | 2007 |
| 7 | Grecce-Italy | 800 | 200 | 1.6 | 2 | ? | 8 | 2012 |
| 8 | Proposed Levitated gas line | 1000 | 1000 | 0.1 | 0.1 | 10 | 25 | - |

Source of 1-7: Wikipedia.

## Load transportation System

Let us take the speed of delivery as equal to $V$ = 30 m/sec, payload capability as 20 - 25 kg per one meter of pipeline in one direction. Then the delivery capability for non-wing containers is 70 kg/s or 20 million tons per year.

That is more than the gas delivery capability (18 million tons per year)! The total load weight suspended under the pipeline of length $L$ = 100 km equals 2500 tons. If a friction coefficient is $f$ = 0.03, the needed thrust is 75 tons and needed power from only friction roller drag is $N_1$ = 22,500 kW (Eq. (10)).

If the air drag coefficient $C_d$ = 0.1, cross section container area $S_c$ = 0.2 m$^2$, the air drag of each single container equals $D_{c1}$ = 2.2 kg, and total drag of 20,000 containers over a length of 100 km is $D_c$ = 44 tons. The need driver power is $N_2$ = 13,200 kW. The total power of transportation system is $N$ = 22500 + 13200 = 35,700 kW. The total thrust force is 77 + 44 = 121 tons.

If $\sigma$ = 200 kg/sq.mm, the cable diameter equals 30 mm.

The suggested delivery system can delivery a weight units (non-wing container) up to 100 kg if a length of container is 5 - 7.5 m.

The pipeline and container delivery capability may be increased by **tens of times** if we will use the wing containers. In this case we are not limited by load levitation capability. The wing container needs just a very light monorail (if any, for guidance rather than support) and a closed-loop thrust cable for drive. That can be used for delivery of water, oil or payload in containers. For example, if our system delivers 4 m$^3$/s, that is equivalent of the conventional river (or a water irrigation canal) having a cross-section area of 20 × 2 m and water speed 0.1 m/s.

If we will use this system for transfer of mechanical energy, we can transfer 35,700 kW for the cable speed at 30 m/sec, and 8 times more by the same cable having a speed of 250 m/sec. Such a conventional power line over 1000 km distances would be more expensive than this entire system.

If the $\alpha$ < 60$^o$ and wing of tube wing device has width of 6 m, our system is stable against a side storm wind 30 - 40 m/s.

The reader can compare the offered installation with known pipelines in Table1 or described in [8]-[16].

**Cost of the load system.** Let us take the cost $100 of 1 container 0.5×0.5×5 m ($C_1$ = $20/m, the cost of the thrust-providing electric motor as $20,000/100km ($C_2$ = $0.2/m), the cost of suspending system $C_3$ = $2/m (light monorail). Then the total cost of material for transport system will be about $22.2/m or $22.2K/km.

The cost of 1 km of 1 line railway is about $(1-2) Million/km, (without engines and tank cars). That means the railway is ~ 400 times more expensive then the offered air load transport system.

**Note about the proposed Dead Sea air water system**. We can use the offered system for delivery of water from the Red or Mediterranean Seas to the Dead Sea. If we will use the wing containers we can deliver up 100 million cubic meters water per year if the diameter of the gas tube is 10 m. That is enough to stop the decline of the Dead Sea if 90% of the Dead Sea surface (except along the coasts where tourists go) is covered by plastic surface film to stop evaporation, not counting on any influx of water from the Jordan River. The preferred option will probably be a gas tube diameter of 30 m (Delivery capacity is 900 M/cubic meters) and omitting covering the Dead Sea surface for aesthetic reasons. The cost of this proposed 30-meter diameter delivery system, if the first one built, will be about 200-700 million dollars (or up to 25 times cheaper than the well-known conventional project Red-Dead Sea (canal) ($5 billion USD.)).  Eventually the cost of a second such pipeline could be half this, and the Dead Sea could in principle be raised to any desired level.

## Acknowledgement

The author wishes to acknowledge Joseph Friedlander for correcting the author's English and for useful technical suggestions.